\documentclass[twocolumn,trackchanges]{aastex631}

\usepackage{booktabs}
\usepackage{graphicx}
\usepackage{amssymb}

\usepackage{multirow}
\usepackage{array}
\usepackage{arydshln}
\begin{document}

\title{Neural Posterior Estimation for UHECR source inference from 3D propagation simulations}

\author[0009-0004-1555-8004]{Nadine Bourriche}
\affiliation{Max Planck Institute for Physics \\
Boltzmannstra\ss e 8, 85748 Garching, Germany}
\affiliation{Technical University of Munich \\ 
James-Franck-Stra\ss e 1, 85748 Garching, Germany}

\author[0000-0002-1153-2139]{Francesca Capel}
\affiliation{Max Planck Institute for Physics \\
Boltzmannstra\ss e 8, 85748 Garching, Germany}

\author[0000-0001-9111-4916]{Nicole Hartman}
\affiliation{Technical University of Munich \\ 
James-Franck-Stra\ss e 1, 85748 Garching, Germany}
\affiliation{ORIGINS Data Science Laboratory \\ 
Boltzmanstra\ss e 2, 85748 Garching, Germany}

\begin{abstract}

The identification of ultra-high energy cosmic ray sources is one of the open challenges of high-energy astrophysics. As charged particles travel through the Universe, they are deflected by extragalactic magnetic fields and lose energy through interactions with background radiation, making source inference highly non-trivial. Existing approaches either rely on simplified propagation models or on computationally prohibitive Monte Carlo methods. Here we present a simulation-based inference framework trained on three-dimensional \texttt{CRPropa~3} propagation simulations that produces calibrated posterior distributions over source energy, distance, direction, and primary composition for individual UHECR events. The model combines a Deep Set encoder, handling the variable number of detected secondary particles, with a normalizing flow, and is trained on approximately 5 million simulated events covering a broad range of extragalactic magnetic field configurations. Validated on held-out simulations, all source parameters are recovered without systematic bias, with directional parameters best constrained and source distance most uncertain, consistent with the underlying propagation physics. Primary composition classification achieves $\geq$~98.2\% accuracy across all mass groups. This framework provides a scalable and physically interpretable interface between detailed propagation simulations and Bayesian source inference relevant for current UHECR data.

\end{abstract}

\section{Introduction} 
\label{sec:intro}

The identification of ultra-high-energy cosmic ray (UHECR) sources is fundamentally challenged by the complex and stochastic nature of their propagation through the Universe. Charged particles experience deflections in both Galactic (GMF) and extra-Galactic magnetic fields (EGMF), as well as energy losses and interactions, such that the mapping between observed properties and source characteristics is highly non-trivial. The Monte Carlo simulation framework \texttt{CRPropa~3} has been developed to provide a detailed description of the forward model of UHECR propagation from sources to observations, including 4D trajectories in space and time \citep{Batista:2022pd}. However, this framework relies on computationally expensive Monte Carlo simulations due to the large volume of the Universe that must be simulated, and the relatively small target that must be hit for UHECRs to be detected on Earth. A targeting mechanism has been developed that offers significant speed-ups for specific scenarios, but the learning algorithm must be re-trained to accommodate different simulation setups while remaining unbiased, again increasing the computational cost when exploring a range of source scenarios \citep{vanVliet:2019ps}. 

As a result, analyses targeting inference of source properties often adopt simplifying assumptions or reduced physical model. For instance, many assume rectilinear propagation of UHECRs, effectively neglecting magnetic deflections \citep{guido2021combined-340}, or restrict the source distribution to a statistically isotropic, homogeneous ensemble rather than tracing individual candidate sources \citep{Heinze:2019ls, muzio2019progress-3e5}. While one-dimensional approaches such as \texttt{SimProp} \citep{Aloisio:2017dg}, \texttt{PriNCe} \citep{Heinze:2019ls}, and \texttt{CRISP} \citep{Morejon:2026pg} provide a more efficient description of UHECR propagation, they rely on simplified treatments of magnetic deflections that can become limiting for source inference. In particular, these approaches typically decouple particle propagation from magnetic deflections or treat the latter in an approximate manner. Such assumptions are generally motivated by the large uncertainties in magnetic field modelling and the limited constraining power of current data sets. However, with ongoing experimental upgrades and future observatories expected to deliver increasingly informative datasets \citep{coleman:2022gh}, in particular the improved measurements of the UHECR composition promised by the AugerPrime upgrade \citep{Castellina:2019jd}, and the larger statistics with the development of TAx4 upgrade \citep{Abbasi:2021pd}, there is a growing need for inference methods that can fully incorporate realistic propagation physics and exploit the improved sensitivity to source properties.

In previous work, we developed an inference framework based on Approximate Bayesian Computation to constrain the origins of individual UHECR events using \texttt{CRPropa~3} simulations \citep{Bourriche:2023od,Bourriche_2026}. While this approach demonstrated the potential of likelihood-free methods in this context, it is fundamentally limited by the need to perform large numbers of forward simulations, making it computationally expensive and difficult to scale to more complex models or larger datasets. Recent advances in simulation-based inference (SBI, see \citealt{Cranmer:2020dg,
deistler2025simulationbasedinferencepracticalguide} for recent reviews), in particular neural posterior estimation, provide a promising alternative by enabling amortized inference: a neural network is trained on simulated data to learn the mapping between observations and model parameters, allowing for rapid posterior evaluation without further simulations. Applications of SBI in the UHECR field have shown strong potential, but have so far focused on one-dimensional propagation models \citep{Bister:2022th}, different inference problems such as air shower reconstruction \citep{Marcias:2026ld}, or on related problems in neutrino astronomy \citep{heyer2026event-a45}. 

In this work, we develop a neural posterior estimation framework trained on \texttt{CRPropa~3} simulations, enabling efficient and scalable inference of the origins of individual UHECR events at the highest energies from observed particle data. More broadly, this approach provides a flexible interface between detailed propagation modelling and statistical inference, with potential applications to larger UHECR sample sizes or related inference problems in the UHECR community. We describe the set up of our \texttt{CRPropa~3} simulations and generation of the training data set in Section~\ref{sec:simset} before describing the implementation of our simulation-based inference framework in Section~\ref{sec:methods}. We then validate our approach and demonstrate the constraining power of our model by testing the method on simulated events in Section~\ref{sec:val}, discussing our findings in Section~\ref{sec:disc} and concluding in Section~\ref{sec:conclusion}.

\section{Physical model} 
\label{sec:simset}

To model the propagation of UHECRs, we use \texttt{CRPropa~3} to perform 3D simulations. We include all relevant particle interactions: photo-pion production, photo-disintegration, electron-pair production, as well as nuclear decay and adiabatic losses.

\begin{table}[t!]
\centering
\caption{Free parameters and their corresponding prior assumptions. The EGMF parameters, $B_\mathrm{rms}$ and $L_c$, are used as conditioning variables for the flow, while the $E_\mathrm{src}$,$D_\mathrm{src}$ and direction $\hat{p}_{i,\mathrm{src}}$ are predicted by the flow, as described in Section~\ref{subsec:preprocess}.
\label{tab:priors}}
\begin{tabular}{llll}
\toprule
\textbf{Parameter} & \textbf{Prior dist.} & \textbf{Range} & \textbf{Unit} \\
\midrule
$B_\mathrm{rms}$ & log uniform & $[0.1,\,10]$ & nG \\
$L_c$ & log uniform & $[60,\,1000]$ & kpc \\
\cdashline{1-4}
$E_\mathrm{src}$ & $\propto E_\mathrm{src}^{-1}$ & $[100,\,1000]$ & EeV \\ 
$D_\mathrm{src}$ & $\propto D_\mathrm{src}^3$ & $[1,50]$ & Mpc \\
$-\hat{p}_{x,\mathrm{src}}$ & \multirow{3}{2.4cm}{uniform on a sphere} & $[-1,1]$ &  \\
$-\hat{p}_{y,\mathrm{src}}$ &                                      & $[-1,1]$ &  \\
$-\hat{p}_{z,\mathrm{src}}$ &                                      & $[-1,1]$ &  \\
\bottomrule
\end{tabular}
\end{table}

All prior choices described here are made for the purposes of demonstrating the framework, and are not fundamental to it: the injected composition, energy range, source distance distribution, and magnetic field priors can easily be altered to target different astrophysical scenarios without architectural changes to the model.

The geometry is designed to maximize the available training data. We place a point source inside a shell-like detector, with source positions drawn uniformly on the sphere at each iteration. The detector distance from the source is a free parameter sampled uniformly within a spherical volume, with distances in the range $[1,50]$\,Mpc, motivated by the expectation that sources of extreme energy events are likely located in the local cosmic web \citep{Kotera2011}.

Source positions are drawn uniformly on the sky at each iteration, reflecting our current lack of established anisotropy in the UHECR arrival distribution at the energies considered and providing uniform coverage of all sky directions for the neural network’s training.

The injected energy spectrum follows a power-law prior such that $\mathrm{d}N/\mathrm{d}E_\mathrm{src} \propto E_\mathrm{src}^{-1}$. This value is the best-fit spectral index obtained by the combined composition and spectrum fit of the Pierre Auger data \citep{Aab2017}. The minimum energy at the source is set to be 100\,EeV, which is driven by our interest in using our model to study the extreme end of the UHECR spectrum. For these events, individual data-driven analysis is well motivated due to the fact that magnetic deflections are typically smaller (while rigidity dependent) and these particles must originate from relatively nearby given the expected interaction loss lengths. The maximum energy is chosen to be 1\,ZeV to explore the impact of trans-GZK energies \citep{Griesen1966}, where the suppression of the flux makes the identification of individual sources particularly compelling and where recently detected events such as Amaterasu \citep{collaboration2023extremely-240} motivate dedicated inference tools \citep{Bourriche_2026}. 

The composition at injection is sampled from the representative set {H, He, N, Si, Fe}, spanning light, intermediate, and heavy mass groups broadly consistent with the composition inferred from air shower measurements \citep{Aab2017}. Equal weighting across compositions is a conservative, uninformative choice that ensures the classifier is trained on representatives of light, medium and heavy mass groups, it does not imply that all compositions are equally abundant at the source.

The EGMF remains relatively poorly constrained, so we adopt broad, physically motivated priors that will not dominate the posterior, given the expectations for the EGMF of the local environment of the Milky Way and nearby Galaxies considered here (e.g.~\citealt{10.1093/mnras/stx3354,Locatelli:2021se}). Moreover, to reflect the large uncertainties while satisfying this constraint, we choose log-uniform priors over the EGMF parameters such that ${B_\mathrm{rms}\sim \log{\mathrm{U}}[0.1,\,10]}$\,nG and ${L_\mathrm{c} \sim \log{\mathrm{U}}[60,\,1000]}$\,kpc, where $B_\mathrm{rms}$ is the strength of the EGMF, and $L_c$ is the global coherence length. 
These values are further constrained by the relation ${\left\langle B_\mathrm{rms}^2 L_\mathrm{c} \right\rangle^{1/2} \lesssim 10^{-8} \, \text{G} \, \text{Mpc}^{1/2}}$ \citep{Kotera2011}. 
We model the EGMF as a Gaussian random field with a Kolmogorov turbulence spectrum on a regular 3D grid with a spacing of 25\,kpc. 

We summarize the free parameters and prior choices in Table~\ref{tab:priors}. These priors are fixed and used throughout this work.

\section{Methods} \label{sec:methods}

\begin{figure*}
    \centering
    \includegraphics[width=1\linewidth]{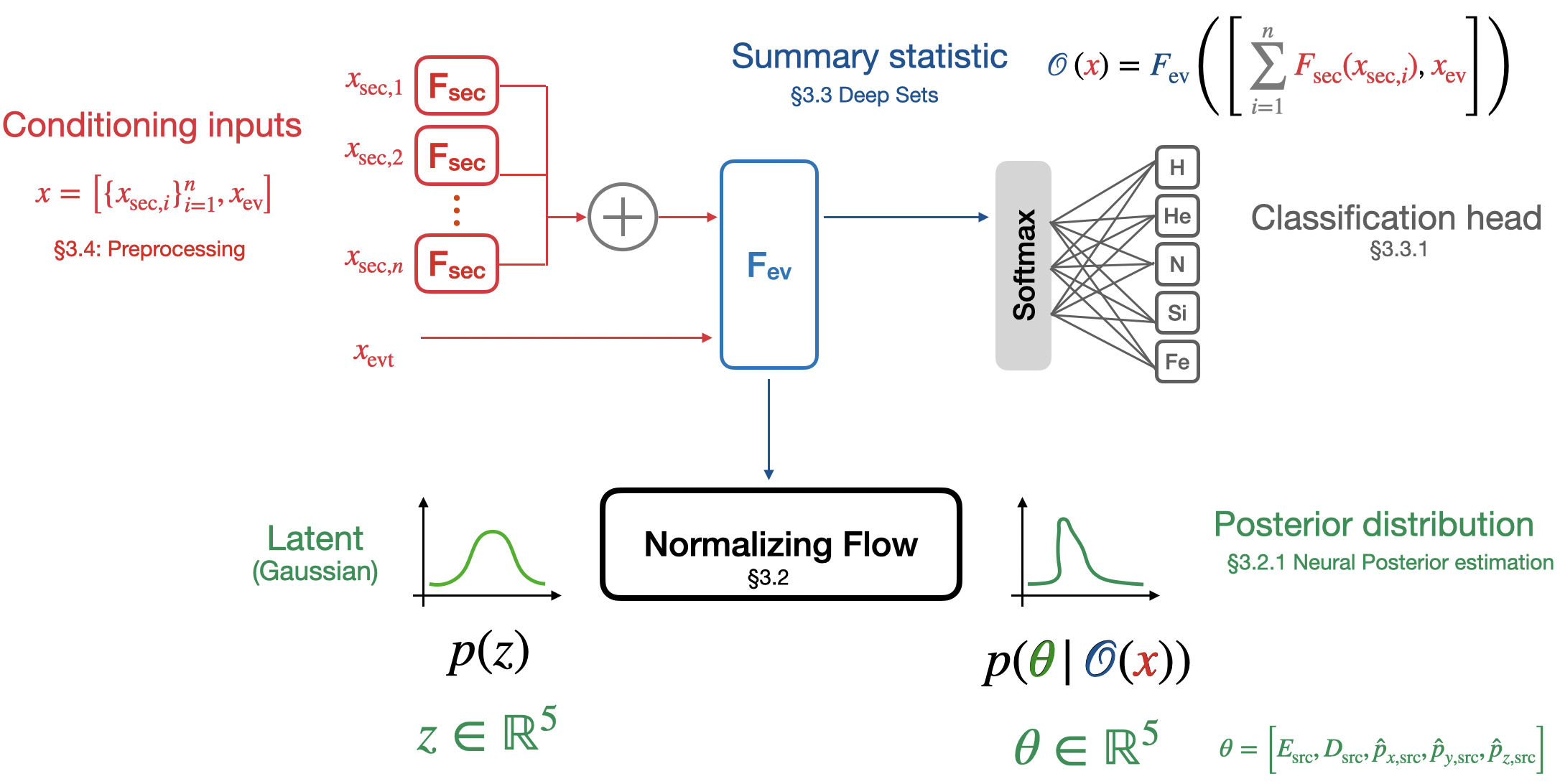}
    \caption{Method for UHECR source estimation. 
    We process the secondaries $\{x_{\mathrm{sec},i}\}_{i=1}^n$ with a Deep Set to flexibly handle the variable cardinality of the secondaries. Inputs $x_{\mathrm{sec},i}$ are processed through per-particle networks $F_\mathrm{sec}$. The latent space $h_\mathrm{sec}$ is concatenated onto the event level inputs $x_\mathrm{ev}$ and correlations are modelled with a per-event neural network, $F_\mathrm{ev}$. The latent space output from $F_\mathrm{ev}$ is then passes to two downstream tasks: (1) The classification head which classifies the source particle type (right), and (2) a conditioning vector for the normalizing flow used for neural posterior estimation (bottom). 
    \label{fig:method}}
\end{figure*}

\subsection{Simulation-based inference}

In the natural sciences, our understanding of the laws of nature is encoded in our forward model or simulator, which provides samples from a stochastic process, \mbox{$x \sim p(x | \theta)$}.
Understanding what parameters $\theta$ might have most likely generated a set of data $x$ requires solving an inverse problem, $p(\theta | x)$.
SBI methods use simulators to infer the optimal parameters given the data \citep{lueckmann2017flexiblestatisticalinferencemechanistic,papamakarios2018fastepsilonfreeinferencesimulation,greenberg2019automaticposteriortransformationlikelihoodfree,Cranmer:2020dg,deistler2025simulationbasedinferencepracticalguide}. 
In previous work \citep{Bourriche_2026}, we used ABC to understand the origin of the Amaterasu particle. The simulator runs multiple times (for different settings for $\{\theta_i\}_{i=1}^{N}$) and saves the $\theta_i$ that produces an observed $x$ within some threshold $\epsilon$ of $\theta$. These saved $\theta_i$ are then used as samples from the posterior distribution $p(\theta |x)$.
With sufficient compute, as $\varepsilon\rightarrow 0$ the ABC method provides accurate posterior samples (assuming sufficient summary statistics).
However, taking $\varepsilon \rightarrow 0$ is enormously computationally expensive: most proposed samples $\theta_i$ are discarded, a challenge that worsens as the dimensionality of the summary statistic increases, the so-called `curse of dimensionality'.

Advancements in generative modeling provide new techniques for learning from the posterior with a neural network. High fidelity simulators can provide training data for data-hungry machine learning models. 
Neural posterior estimation (NPE) directly learns an approximation to the posterior $q_\phi(\theta \mid x)$ from pairs of samples $(\theta, x)$, where $\phi$ parametrizes a neural density estimator. %

Fig.~\ref{fig:method} shows an overview of our method. 
In the following we describe: (1) the neural network we use for NPE; (2) the deep set used for the data compression neural network; and (3) the technical details for our implementation of NPE for \texttt{CRPropa~3}.

\subsection{Normalizing Flows}
\label{subsec:flows}
Normalizing flows are an attractive architecture for SBI as they are designed to both evaluate the likelihood exactly and quickly generate samples, (see \citet{papamakarios2019neuraldensityestimationlikelihoodfree,papamakarios2021normalizingflowsprobabilisticmodeling} for an introduction and review). We summarize the key ideas below.

A normalizing flow builds a 
transformation \mbox{$f_\phi: z \mapsto \theta$} between two distributions $q_Z(z)$ and $q_\phi(z)$ on $\mathbb{R}^d$, where $q_Z(z)$ is the base distribution (usually Gaussian) that is easy to evaluate the likelihood and sample from.
By the conservation of probability, these distributions are related as:
\begin{equation}
    q_\phi(\theta) = \left|\frac{df_\phi^{-1}}{d\theta}\right| q_Z(f_\phi^{-1}(x)) = \left|\frac{df_\phi}{d\theta}\right|^{-1} q_Z(f_\phi^{-1}(\theta)).
    \label{eq:consprob}
\end{equation}
Neural networks (NNs) are powerful function approximators and are a natural choice to model $f_\phi$. However, Eq.~\ref{eq:consprob} sets constraints on a NN that could model $f_\phi$: the $f_\phi$ network must be invertible with a tractable Jacobian\footnote{For $\theta \in \mathbb{R}^d$, computing the Jacobian naively using automatic differentiation with an unstructured transformation is $\mathcal{O}(d^3)$, too expensive to use in a training optimization.}, $\left|\frac{df_\phi}{dx}\right|$.
A normalizing flow builds up the complicated transformation $f_\phi$ by a series of simpler transformations $\{f_{\phi,i} \}_{i=1}^L$ so that $f = f_{\phi,L} \circ \cdots f_{\phi,2} \circ f_{\phi,1}$. 
If each of the $f_{i,\phi}$ is invertible with a tractable Jacobian, then $f_\phi$ is as well.
The research into normalizing flow architectures in the past years involve different methods to restrict the form of $f_{\phi,i}$ to satisfy these conditions.
By using a sequence of $f_{\phi,i}$, each is only a small transformation, and the $q_Z(z)$ distribution gradually morphs or flows one distribution into the other, ergo the name, `normalizing flows'.

The two directions of $f_\phi$ fulfill our two desiderata:
\begin{itemize}
    \item \emph{Training:} Given a data sample $\theta$, applying $f_\phi^{-1}$ lets us evaluate the probability $q_\phi(\theta)$ using Eq.~\ref{eq:consprob}. The parameters $\phi$ of $f_\phi$ are then optimized by maximum likelihood, or equivalently, by minimizing the negative log-likelihood as a loss function.
    \item \emph{Sampling:} Given a noise sample $z \sim q_Z(z)$ from the base distribution, we can apply the flow $f_\phi(z)$ to obtain a sample from $q_\phi(\theta)$.
\end{itemize}

Due to its flexibility and easy training, we use the Rational Quadratic Neural Spline Flow (RQ-NSF) from \cite{durkan2019neuralsplineflows}, a common model for SBI applications \citep{vandegar2021neuralempiricalbayessource,
dax2021realtime-43b,heinrich2024hierarchicalneuralsimulationbasedinference,
kofler2025flexiblegravitationalwaveparameterestimation}.
In a RQ-NSF, the transform $f_{\phi, i}$ is a spline described by the ratio of two quadratic polynomials. 
The networks (with trainable parameters $\phi$) predict the knot locations and derivatives for the spline.
Both $f^{-1}_{\phi, i}$ and its Jacobian can be inverted in $\mathcal{O}(d^2)$.

The hyperparameters are: the number of knots $K$, the domain for the spline (parametrized as [-B,B]),  the number of layers of the flow, $L$, the hidden dimension, $H$ for the NNs, and the number of residual blocks used to predict these NNs. To mitigate overfitting, the dropout fraction \citep{JMLR:v15:srivastava14a} and L2 regularization are additional hyperparameters.
An identity transformation is used to extrapolate when $\theta_i$ falls outside of the domain of the spline.

\subsubsection{Neural Posterior Estimation}

In NPE we don't only estimate a density $q_\phi(\theta)$, but a rather a conditional density, $q_\phi(\theta|x)$. To modify the above presentation for conditional density estimation, extra conditioning variables, $x$, are included as  additional inputs to parametrize the flow NNs. As observations $x$ are often much higher dimensional than the $\theta$ dimensionality, the data $x$ is often first compressed into a \emph{summary statistic} $\mathcal{O}(x)$ pass as an input to condition the flow.
This $\mathcal{O}$ can be learned with an additional NN trained end-to-end with the flow \citep{dax2021realtime-43b}.

For training data, samples are drawn from the joint distribution $(\theta, x) \sim p(\theta, x) = p(x \mid \theta)\, p(\theta)$ through hierarchical sampling.  $\theta \sim p(\theta)$ is drawn from the prior, and $x \sim p(x \mid \theta)$ is obtained from the simulator. Thus, the choice of prior is implicitly encoded in the training dataset.

\subsection{Deep set encoder}
\label{subsec:deepset}

The last challenge comes from the structure of the data: our observations are $x = \left[\{ x_{\mathrm{sec},i} \}_{i=1}^n,  x_\mathrm{ev} \right]$, where $x_{\mathrm{sec},i}$ are the features per secondary, and $x_\mathrm{ev}$ is the event level feature. We need to compress the variable length $x_\mathrm{sec}$ to create to a fixed dimensional conditioning vector.
We're inspired by \citet{heinrich2024hierarchicalneuralsimulationbasedinference} use of a deep set encoder \citep{{zaheer2018deepsets}} to compress data examples with varying cardinality for SBI. 
Deep set's are also common for classification in particle physics \citep{Komiske_2019,ATLAS_DIPS}.

The deep set combines a per-secondary network $F_\mathrm{sec}$ with an event level network $F_\mathrm{ev}$ to obtain a fixed-dimensional, permutation-invariant representation of the variable length set of detected secondaries.
For each detected secondary $i$, we partition the features: \mbox{$x_{\mathrm{sec},i} = (\tilde{x}_{sec,i}, \mathrm{ID}_i)$}, where $\tilde{x}_\mathrm{sec,i}$ contains the continuous features \mbox{$\tilde{x}_{\mathrm{sec},i} = [\log_{10}E_i,\,-\hat{p}_{x,i},-\hat{p}_{y,i},\,-\hat{p}_{z,i}]$}, and particle ID is an integer. 
We use a learned 16 dimensional embedding $\mathrm{Emb}(\mathrm{ID}_i)$ for the categorical particle ID, and concatenate it to the continuous features.
Applying the $F_\mathrm{sec}$ network yields a latent representation for each secondary \mbox{$h_i = F_\mathrm{sec}([\tilde{x}_{\mathrm{sec},i},\mathrm{Emb}(\mathrm{ID}_i)]) \in \mathbb{R}^{h_\mathrm{sec}}$}. 
To accommodate variable multiplicities, each event is padded to a fixed maximum number of secondaries $N_{\max}$ and a binary mask is applied such that padded entries do not contribute to the aggregation. 

We compute an event summary by summarizing over these secondary latents using masked mean pooling, \mbox{$\bar{h} = \frac{1}{n}\sum_{i=1}^{n} h_i \in \mathbb{R}^{h_\mathrm{sec}}$}. 
We have other event features $x_\mathrm{ev} = (L_c,B_\mathrm{rms},log_{10}(\mathrm{flux_{sum}}),n/n_{\max})$ where $L_c$ and $B_\mathrm{rms}$ describe the EGMF, $\mathrm{flux_{sum}}$ is the sum of the energy from the secondaries in the event, and $n/n_{\max}$ is the occupancy fraction, added because mean pooling suppresses explicit information about the number of secondaries.
The event level network $F_\mathrm{ev}$ network maps the concatenated vector $[\bar{h},x_{\mathrm{ev}}]$ to a context vector $\mathcal{O} = F_\mathrm{ev}([\bar{h},x_{\mathrm{ev}}]) \in \mathbb{R}^{h_\mathrm{ev}}$.
This context conditions the normalizing flow posterior.


\subsubsection{Source classification head}

Finally for the particle source classification, the cross entropy loss provides a distribution over the probabilities.
The deep set's context $\mathcal{O}(x)$ is also fed to a source classification head which is a linear + Softmax layer to predict the primary label $\mathrm{ID}_0$.

\subsection{Inputs and preprocessing}
\label{subsec:preprocess}

Each event is described by an event-level feature vector ${x}_\mathrm{ev}$ and a per-secondary feature vector $x_\mathrm{sec}$, where events with fewer than 80 secondaries are zero padded. Continuous features are standardized using training set moments (mean and standard deviation).

The regression target (free parameters) for each event is $\theta_i = \{\log_{10}E_{\mathrm{src}},\, D_\mathrm{src},\,
-\hat{p}_{x,\mathrm{src}},-\hat{p}_{y,\mathrm{src}},\,-\hat{p}_{z,\mathrm{src}}\}$ which encodes the source energy, the unit vector pointing from the observer towards the source, and a log-transformed source distance. The distance is subtracted from $D_{\mathrm{max}}+1 = 51$\,Mpc so that the transformed variable increases monotonically as the source approaches the observer to aid the network learning. All five components are likewise standardized using training set moments. 

A careful distinction must be made between parameters we infer and those we condition on. Table~\ref{tab:priors} lists all simulation parameters with their prior ranges, but not all appear in the regression target, $\theta$. The EGMF parameters $B_{\mathrm{rms}}$ and $L_c$ instead enter the model as conditioning variables within $x_{\text{ev}}$. This design choice is pragmatic: marginalizing over the EGMF would require the flow to learn a joint posterior $p(\theta, B_{\mathrm{rms}}, L_c \mid x)$, and early experiments showed this demanded substantially more training data for stable calibration. By conditioning on $B_{\mathrm{rms}}$ and $L_c$ instead, the model focuses its representational capacity on the source parameters of primary interest. To perform inference on real data, one can either (i) run the model with fixed EGMF realizations or (ii) marginalize over EGMF uncertainties by averaging predictions across an ensemble of plausible field configurations drawn from their priors. We leave the development of a fully marginalized model that simultaneously infers $B_{\mathrm{rms}}$ and $L_c$ to future work.

In addition, each event carries a discrete label \mbox{$\mathrm{ID}_0 \in \{^1\mathrm{H},\,^4\mathrm{He},\,^{14}\mathrm{N},\,^{28}\mathrm{Si},\,^{56}\mathrm{Fe}\}$}
corresponding to the true primary cosmic-ray composition, which is used to evaluate the classification performance of the model.

We simulate 6~million events, one primary particle per iteration,
where each simulated point draws a new sample from the prior. 
As magnetic deflections cause particles to travel curved trajectories, their total propagated path length can substantially exceed the source distance. We set a deactivation threshold at 70\,Mpc of total path length; particles that propagate beyond this distance without reaching the detector are deactivated. The final dataset contains approximately 5.3~million particles.

\subsection{Hyperparameters and training}
\label{subsec:techdetails}

The hyperparameters of the normalizing flow were selected by evaluating five configurations inspired by the optimal parameters
reported in \cite{durkan2019neuralsplineflows}. 
Each configuration varies the number of coupling layers, the width of the context networks, the batch size, the learning rate, and the dropout fraction.
For all configurations we use the AdamW optimizer, linear tails for the RQ-spline outside the domain, and LU-decomposed invertible linear mixing layers interleaved with the coupling transforms \citep{dinh2017densityestimationusingreal}, following the RQ-NSF (C) architecture of \cite{durkan2019neuralsplineflows}. Each configuration was trained for 50 epochs, and Configuration 3 achieved the lowest validation NLL and was selected for the full training run, corresponding to $L=20$, $H=128$, $K=8$, batch size 256, learning rate $5\times10^{-4}$, and dropout $p_{drop}=0.2$.

For the network $F_\mathrm{sec}$ has 2 layers, implemented as two fully connected linear transformations, each followed by a ReLU activation and LayerNorm \citep{ba2016layernormalization}, with a final dropout layer \citep{JMLR:v15:srivastava14a} for regularization. 
The event-level network $F_\mathrm{ev}$ consists of a single fully connected hidden layer, followed by ReLU and LayerNorm. Both $h_\phi$ and $h_F$ have a latent dimensionality of 128, matching the width of the hidden layers.

The full model is trained by minimizing a combined loss \(\mathcal{L} = \mathcal{L}_{\mathrm{NLL}} + \lambda \mathcal{L}_{\mathrm{CE}}\), where \(\mathcal{L}_{\mathrm{NLL}} = -\mathbb{E}[\log q_{\mathrm{\phi}}(\theta \mid x)]\) is the negative log-likelihood (NLL) of the normalizing flow and \(\mathcal{L}_{\mathrm{CE}}\) is the cross-entropy (CE) loss for the auxiliary ID\(_0\) classification head. We use the AdamW optimizer with learning rate \(5 \times 10^{-4}\) and weight decay \(10^{-4}\), with gradient norm clipping at 5.0. 

Training is performed on 4\,239\,530 events with a batch size of 256, with 20\% of the dataset held out as a validation set that is not seen during optimization. The negative log-likelihood measures how much probability the flow assigns to the true source parameters given the observations: more negative values indicate that the model places higher probability mass near the correct answer. To prevent overfitting, early stopping is applied with a patience of 50 epochs, halting training if the validation NLL does not improve by at least \(10^{-4}\). In practice, training stopped after 230 epochs, reaching a final training NLL of approximately \(-8.33\) and a best validation NLL of \(-9.11\) and a final training CE loss of \(0.0149\) and a best validation CE of \(0.0137\) .

\section{Model evaluation}\label{sec:val}

We perform three diagnostic tests to assess the model's calibration and predictive performance: a prior recovery test, comparison plots of posterior means against true values, and individual posterior corner plots.

We note that although the model regresses the source direction as a Cartesian unit vector, we convert to the more interpretable Galactic coordinates $(g_{\mathrm{lon}}, g_{\mathrm{lat}})$.

The first diagnostic that we perform on the model to check its predictive power is a prior recovery test \citep{talts2018}. 
The conditional flow defines $q_{phi}(\theta|x)$, for a selected set of $N$ validation events $\{x_{i}\}_{i=1}^{N}$ we draw $S$ Monte Carlo samples per event $\theta_{i,1},\theta_{i,1} ... \theta_{i,S} \sim q_{\phi}(\theta|x_{i})$.
We then pool the per event samples which produces an empirical distribution $q_{MC}(\theta)=\frac{1}{N}\sum^N_{i=1}q_{\phi}(\theta|x_i)$ which for a large and representative distribution of $x_{i}$ converges to the model marginal $q_{\phi}(\theta)$. We perform this test by choosing 10\,000 random validation events and sampling 1\,000 times from each of their posteriors.

To quantify the agreement between the recovered and true marginals we adopt the Wasserstein-1 distance, \(W_1\). For two univariate distributions \(a\) and \(b\) it is defined as, 

\begin{equation}
W_1(a,b) = \int_{-\infty}^{\infty} |G_a(x) - G_b(x)|dx,
\end{equation}

where \(G_a\) and \(G_b\) are the respective cumulative distribution functions. We prefer \(W_1\) over alternatives such as the Jensen-Shannon divergence because it is sensitive to both shape differences and systematic shifts between distributions, and because it respects the geometry of the parameter space by assigning larger penalties to mass displaced further from its origin \citep{arjovsky2017, villani2009}. It has been adopted as the standard evaluation metric in analogous flow based inference problems in physics, most notably in gravitational wave parameter estimation \citep{dax2021realtime-43b}. To enable comparison across parameters with different physical scales, we additionally report \(W_1\) normalized by the width of the prior interval for each parameter, \(W_1^{\mathrm{norm}} = W_1\,/\,\Delta\theta \times 100\%\), as shown in Table~\ref{tab:w1}.

\begin{table}[h]
\centering
\caption{Wasserstein-1 distances ($\downarrow$ is better) between true and recovered marginal priors
for all source parameters. Physical \(W_1\) values are expressed in the
native units of each parameter. Normalized values \(W_1^{\mathrm{norm}}\)
are computed as a percentage of the parameter support range.}
\label{tab:w1}
\begin{tabular}{lccc}
\toprule
\textbf{Parameter} & \textbf{Range} & \boldmath\(W_1\) & \boldmath\(W_1^{\mathrm{norm}}\) \\
\midrule
\(g_{\mathrm{lon}}\) & \([-180^{\circ},+180^{\circ}]\) & \(0.40^{\circ}\) & \(0.11\%\) \\
\(g_{\mathrm{lat}}\) & \([-90^{\circ},+90^{\circ}]\) & \(0.24^{\circ}\) & \(0.13\%\) \\
\(D_{\mathrm{src}}\) & \([1,50]~\mathrm{Mpc}\) & \(0.22~\mathrm{Mpc}\) & \(0.44\%\) \\
\(E_{\mathrm{src}}\) & \([100,1000]~\mathrm{EeV}\) & \(3.74~\mathrm{EeV}\) & \(0.42\%\) \\
\bottomrule
\end{tabular}
\end{table}

For reference, a maximally unsuccessful recovery (ie. a model that puts all its probability on a single value at one extreme of the parameter range, compared against a true uniform prior) yields \(W_1^{\mathrm{norm}} = 50\%\) by the properties of the Wasserstein distance \citep{villani2009}. The sub-percent agreement obtained here indicates highly accurate prior recovery. 
This upper bound is exact only for \(g_{\mathrm{lon}}\) which follows a uniform prior. For \(D_{\mathrm{src}}\) and \(E_{\mathrm{src}}\), whose priors are non-uniform, the bound serves as an order of magnitude reference rather than a strict limit.

\begin{figure*}[ht]
  \centering
  \gridline{%
    \fig{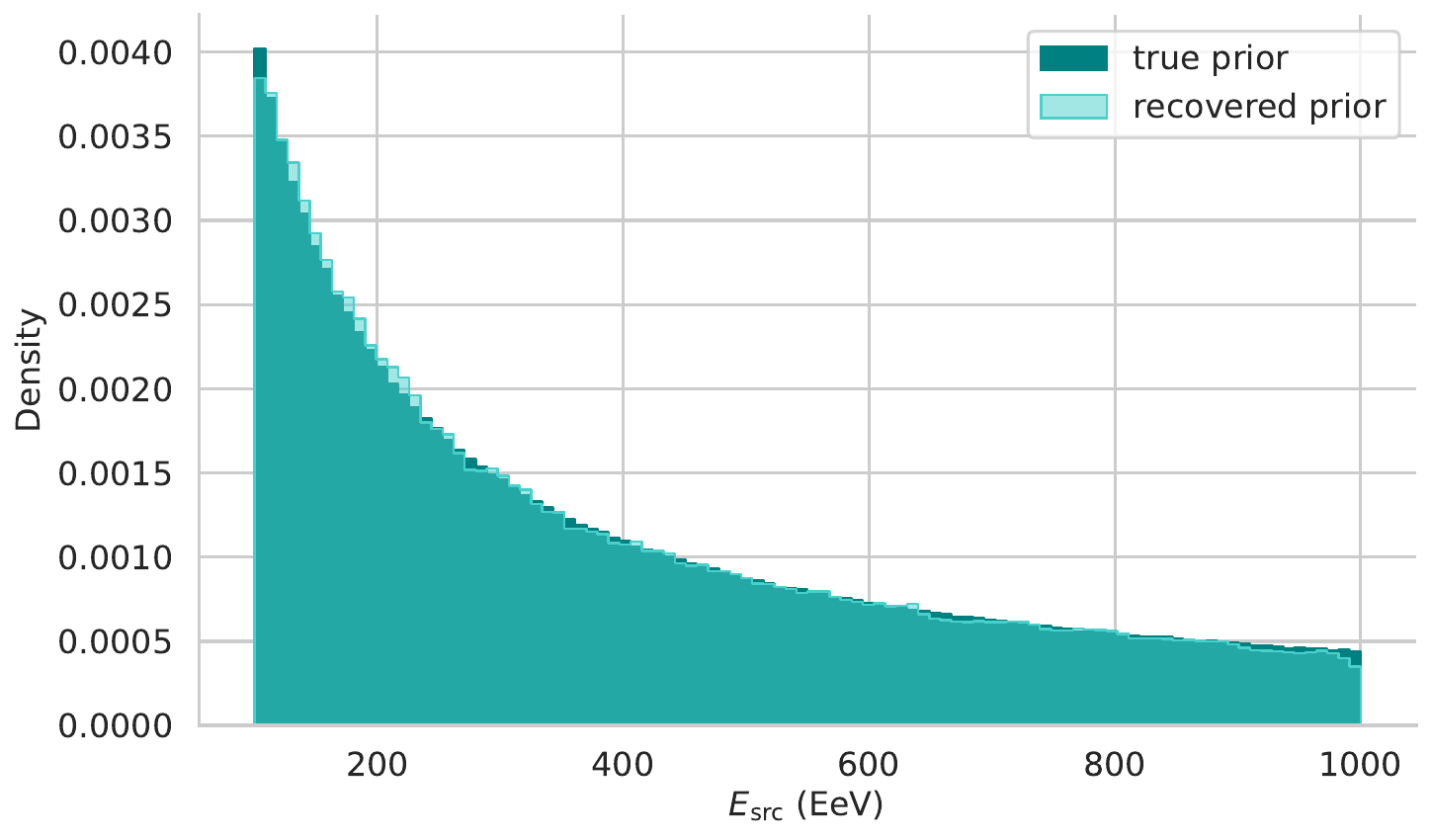}{0.45\textwidth}{(a)}
    \fig{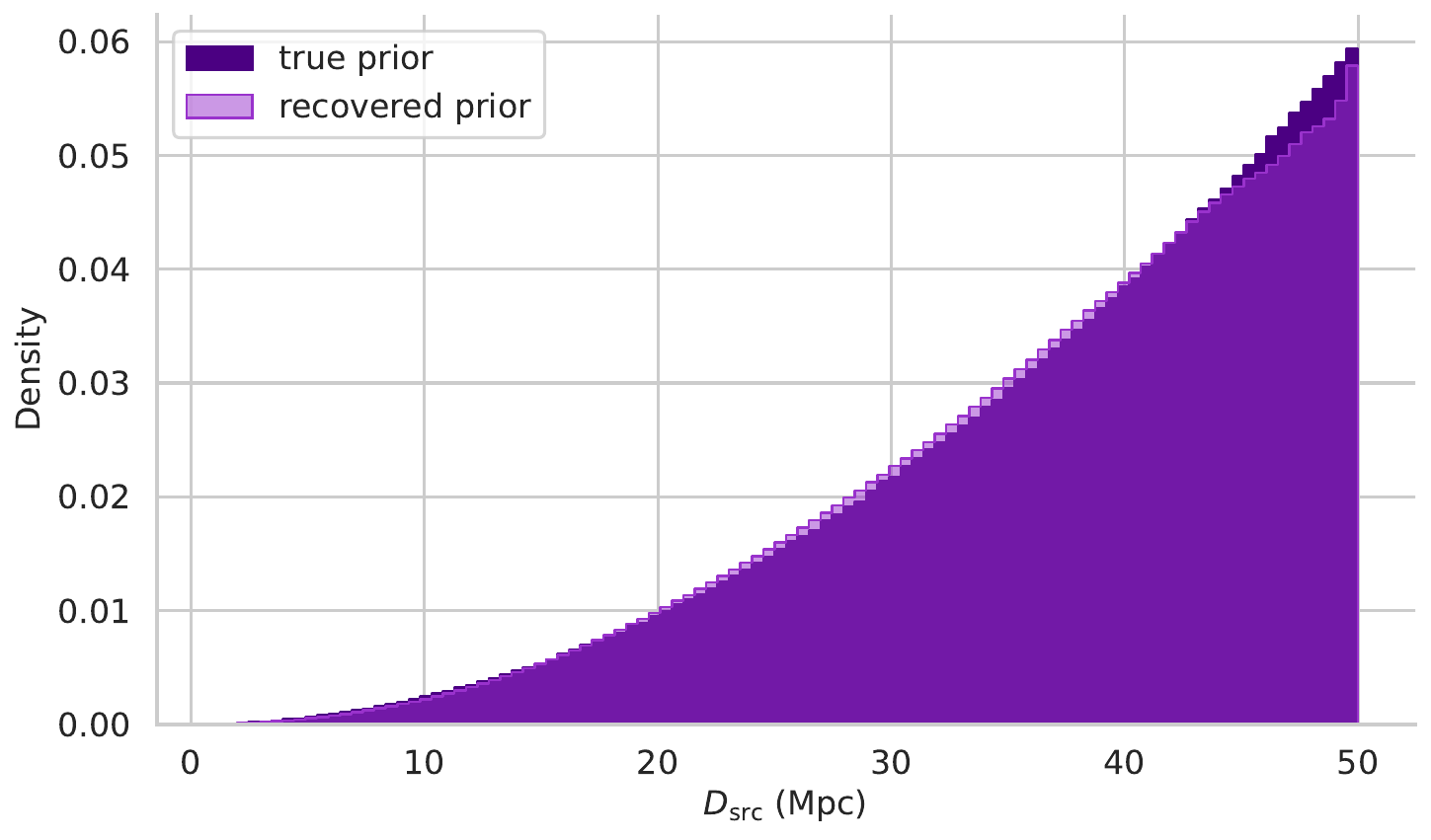}{0.45\textwidth}{(b)}
  }
  \vspace{6pt}
  \gridline{%
    \fig{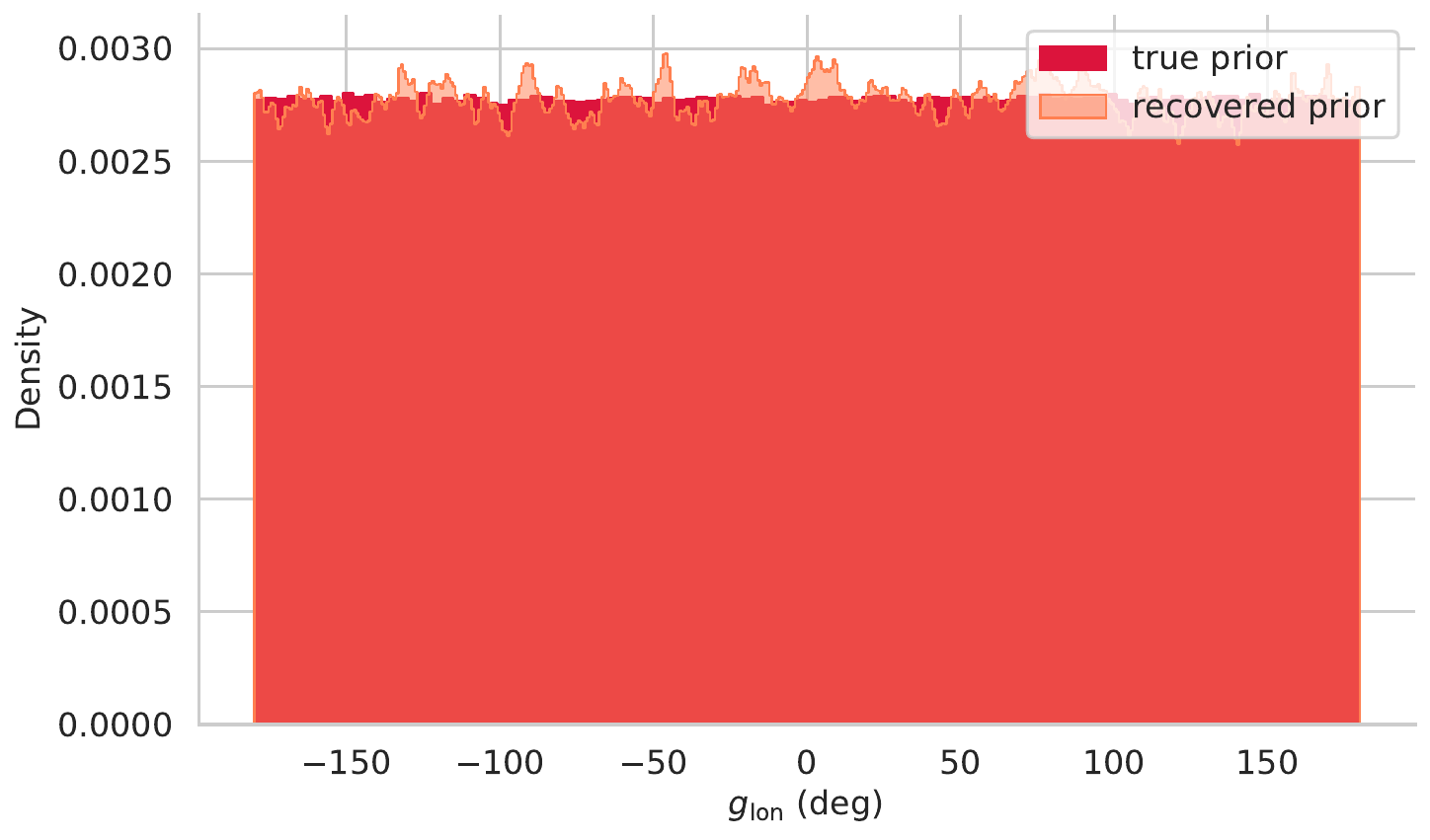}{0.45\textwidth}{(c)}
    \fig{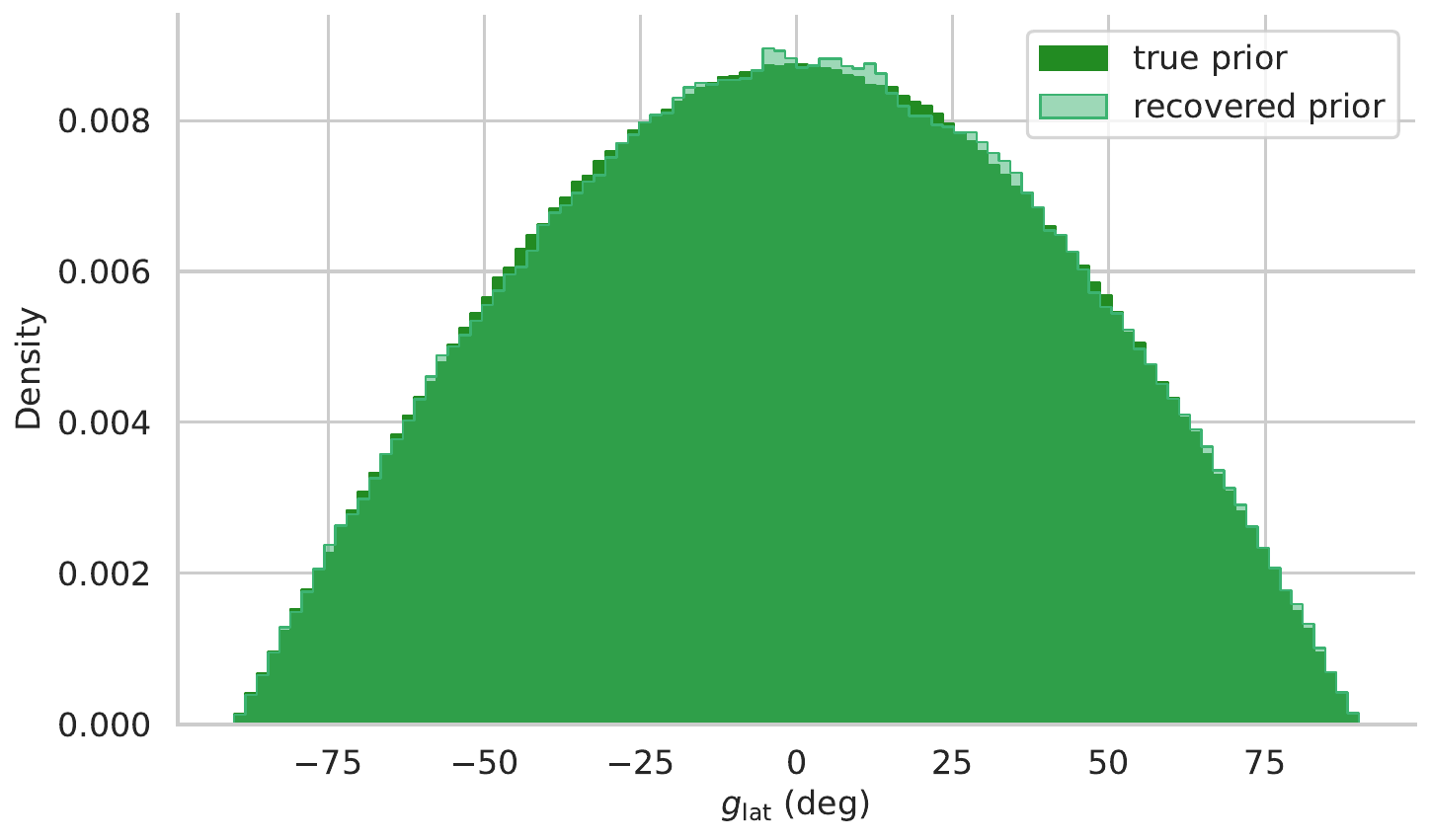}{0.45\textwidth}{(d)}
  }
  \caption{True versus recovered parameter values from the prior
    recovery test. Each panel compares the ground truth value to the
    recovered prior: (a) energy at the source, (b) distance of the
    source, (c) galactic longitude of the source, and (d) galactic latitude of the source.
    \label{fig:prior_test}}
\end{figure*}

In Fig.~\ref{fig:prior_test}, we show the results of the prior recovery test. All four priors are recovered correctly, and the residual differences are small, likely due to Monte Carlo noise from finite sampling. For $D_{\mathrm{src}}$ and $E_{\mathrm{src}}$ we observe slight discrepancies at the extreme low and high values, which is expected because priors that deviate strongly from a normal distribution, being particularly steep and having long tails, are harder for the model to learn. This result is also reflected in the calculated \(W_1^{\mathrm{norm}}\), which are 0.44\% and 0.42\% respectively. While \(g_{\mathrm{lon}}\) and \(g_{\mathrm{lat}}\) are recovered better with \(W_1^{\mathrm{norm}}\) of 0.11\% and 0.13\%, respectively.

\begin{figure}[ht]
    \centering
    \includegraphics[width=\columnwidth]{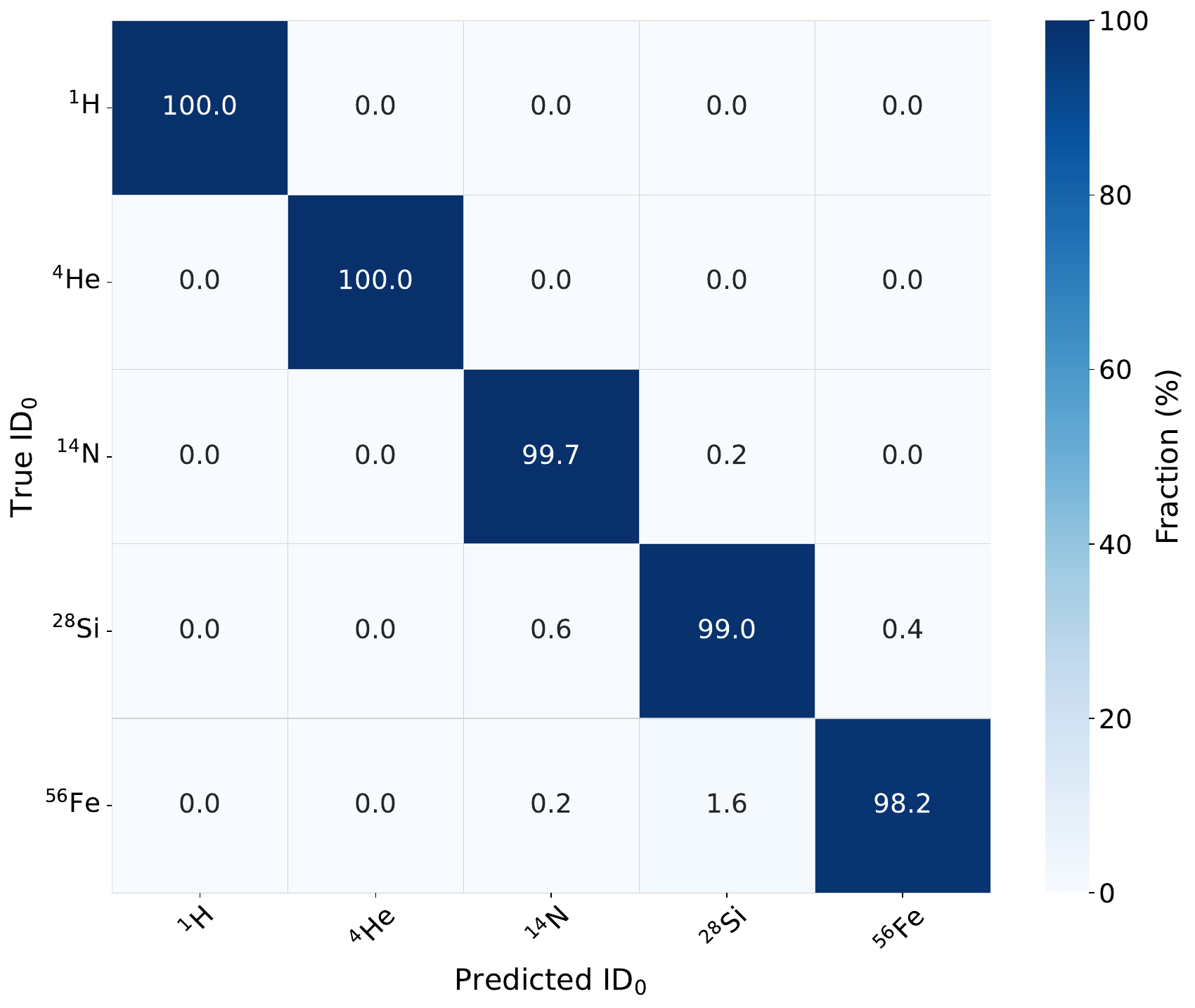}
    \caption{Normalized confusion matrix for primary cosmic-ray composition classification on the full validation set. Rows correspond to true compositions and columns to model predictions, with each entry expressing the fraction of true-class events assigned to the predicted class. The lightest composition, \(^1\mathrm{H}\) and \(^4\mathrm{He}\), are classified with 100\% accuracy. The heavier nuclei \(^{14}\mathrm{N}\), \(^{28}\mathrm{Si}\), and \(^{56}\mathrm{Fe}\) are correctly identified at the 99.7\%, 99.0\%, and 98.2\% level respectively, with the small residual misclassification confined exclusively to adjacent mass groups, consistent with the charge dependent similarity of their propagation signatures.}
    \label{fig:confusion_matrix}
\end{figure}

To assess the power of the source classification head, we present a confusion matrix in Fig. \ref{fig:confusion_matrix}, where we see that the prediction accuracy degrades for heavier nuclei. $^1H$ and $\mathrm{^4He}$ are predicted correctly in 100\% of cases, while $\mathrm{^{14}N}$ is predicted correctly in 99.7\%, $\mathrm{^{28}Si}$ in 99\% of cases, and $\mathrm{^{56}Fe}$ in 98.2\% of cases.

\begin{figure*}
    \centering
    \includegraphics[width=1\linewidth]{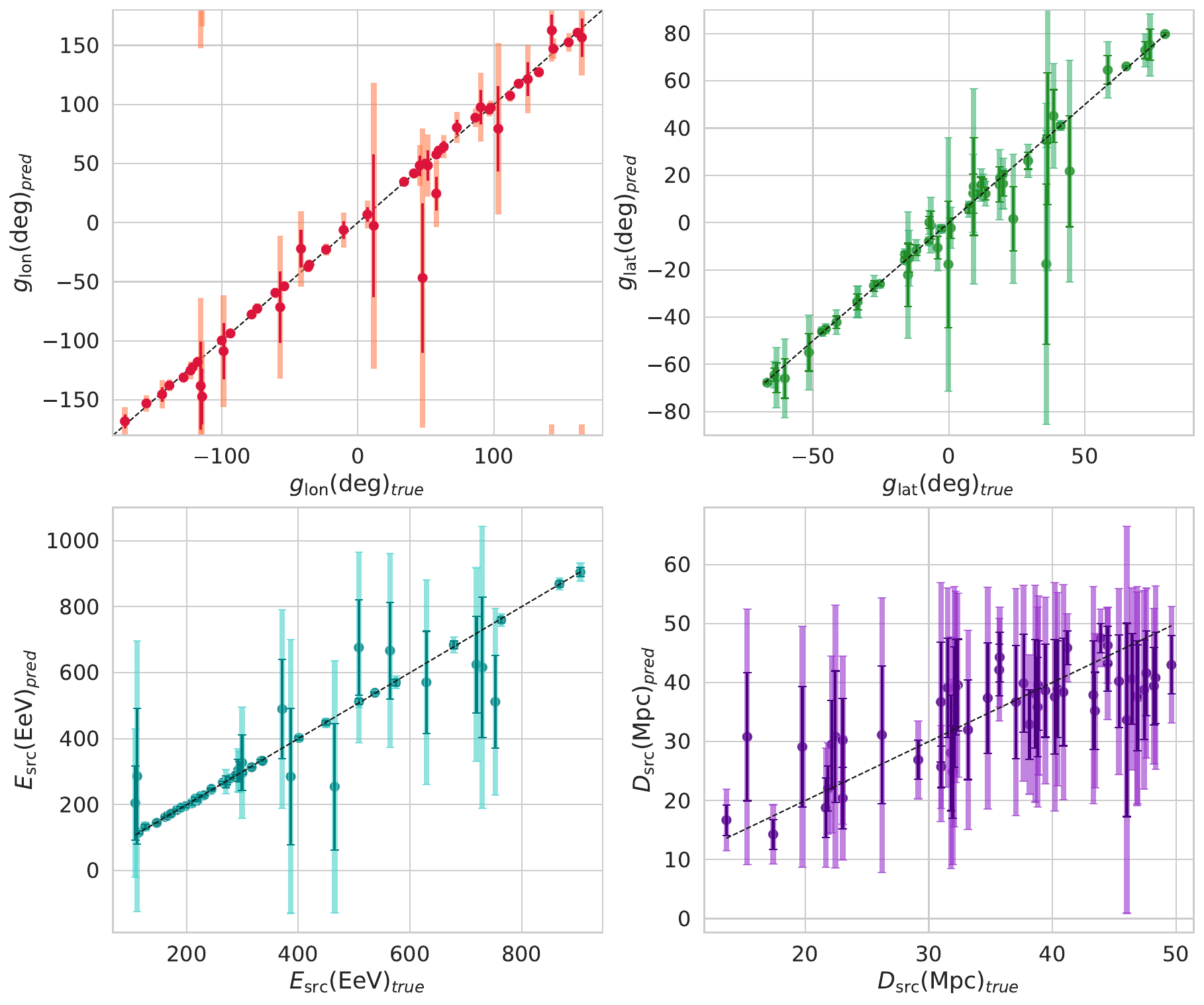}
    \caption{Scatter plots of the true values of 50 randomly picked validation events compared with the predicted values. On the upper panel the results for the source longitude and latitude, in the lower panel the results for the source energy and source distance. In dark red, green, blue and purple the $1\sigma$ error bars and in light red, green, blue and purple the $2\sigma$ error bars.\label{fig:scatter}}
\end{figure*}

The next test comprises a comparison of the posterior means and true values for the other free parameters in our model. We randomly select 50 validation events, draw 10\,000 samples from each marginal posterior, compute the posterior mean and the 68\% and 95\% credible intervals, and then produce true vs. predicted plots for all four parameters. In Fig.~\ref{fig:scatter}, we can see that the means follow the one-to-one relation for all four parameters, indicating overall agreement between the posteriors and the
true values. For the Galactic longitude \(g_{\mathrm{lon}}\), the posterior means lie close to the identity line across the full \([-180^{\circ}, +180^{\circ}]\) range, confirming that the model produces unbiased longitude estimates on average. However, the credible intervals are broad for many events: while the 68\% intervals are moderate in size, the 95\% intervals are large and in some cases span a substantial fraction of the full angular range, reflecting the genuine degeneracy introduced by magnetic deflections.

\begin{figure*}[ht]
    \centering
    \includegraphics[width=1\linewidth]{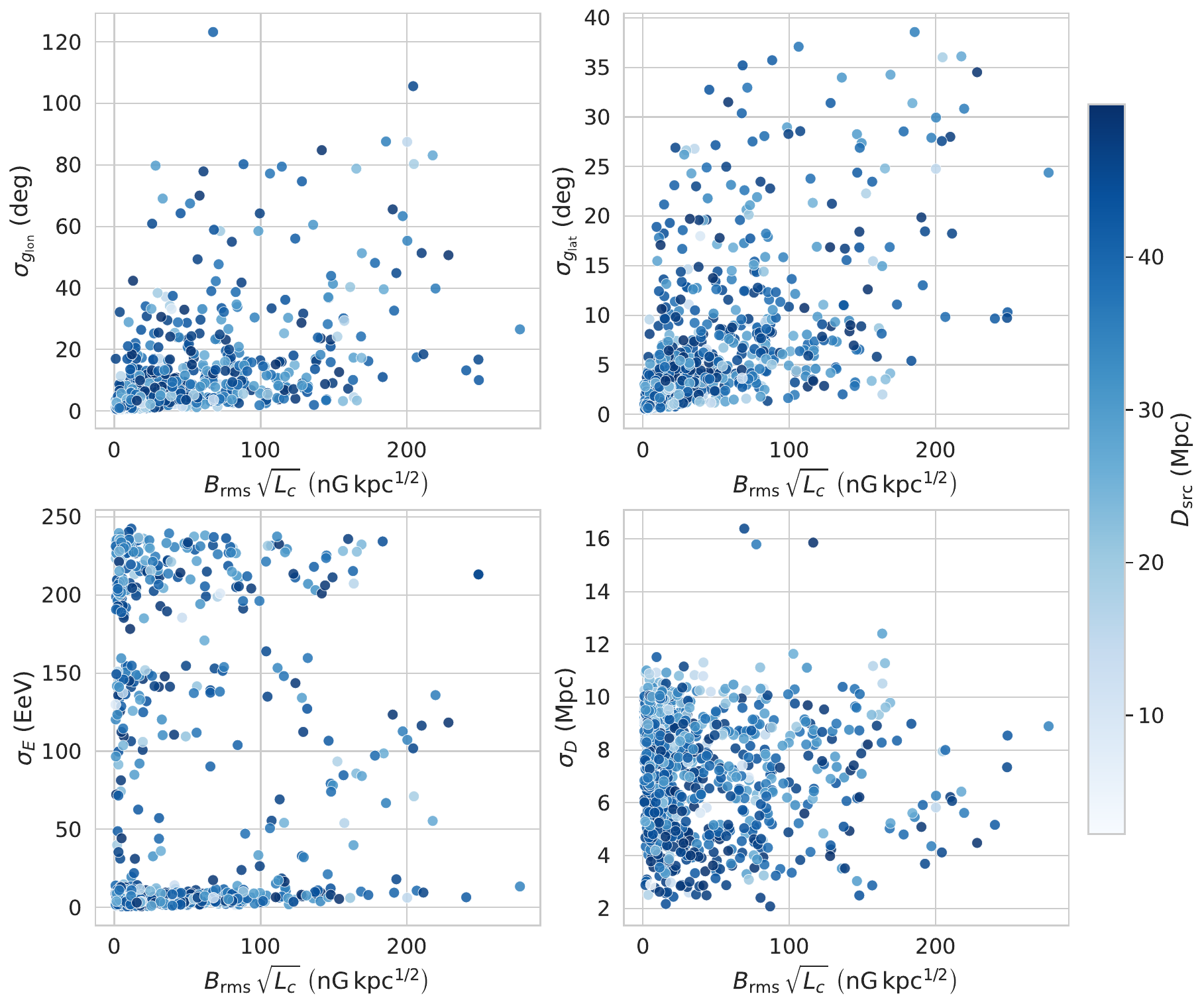}
    \caption{Posterior uncertainty \(\sigma\) for all four source parameters as a function of \(B_\mathrm{rms}\sqrt{L_c}\), with points colored by the true source distance \(D_\mathrm{src}\). Each panel corresponds to one of the four inferred source parameters: Galactic longitude \(\sigma_{g_{lon}}\), Galactic latitude \(\sigma_{g_{lat}}\), source energy \(\sigma_E\), and source distance \(\sigma_D\).}
    \label{fig:sigma_brms}
\end{figure*}

We can investigate this effect further through Fig.~\ref{fig:sigma_brms}, which shows the posterior uncertainty for the four source parameters as a function of $\mathrm{B_{rms}\sqrt{Lc}}$, with the scatter points color coded based on the source distance. We can clearly see a correlation between the strength of the EGMF and the directional uncertainty. 

\begin{figure*}[ht]
    \centering
    \includegraphics[width=1\linewidth]{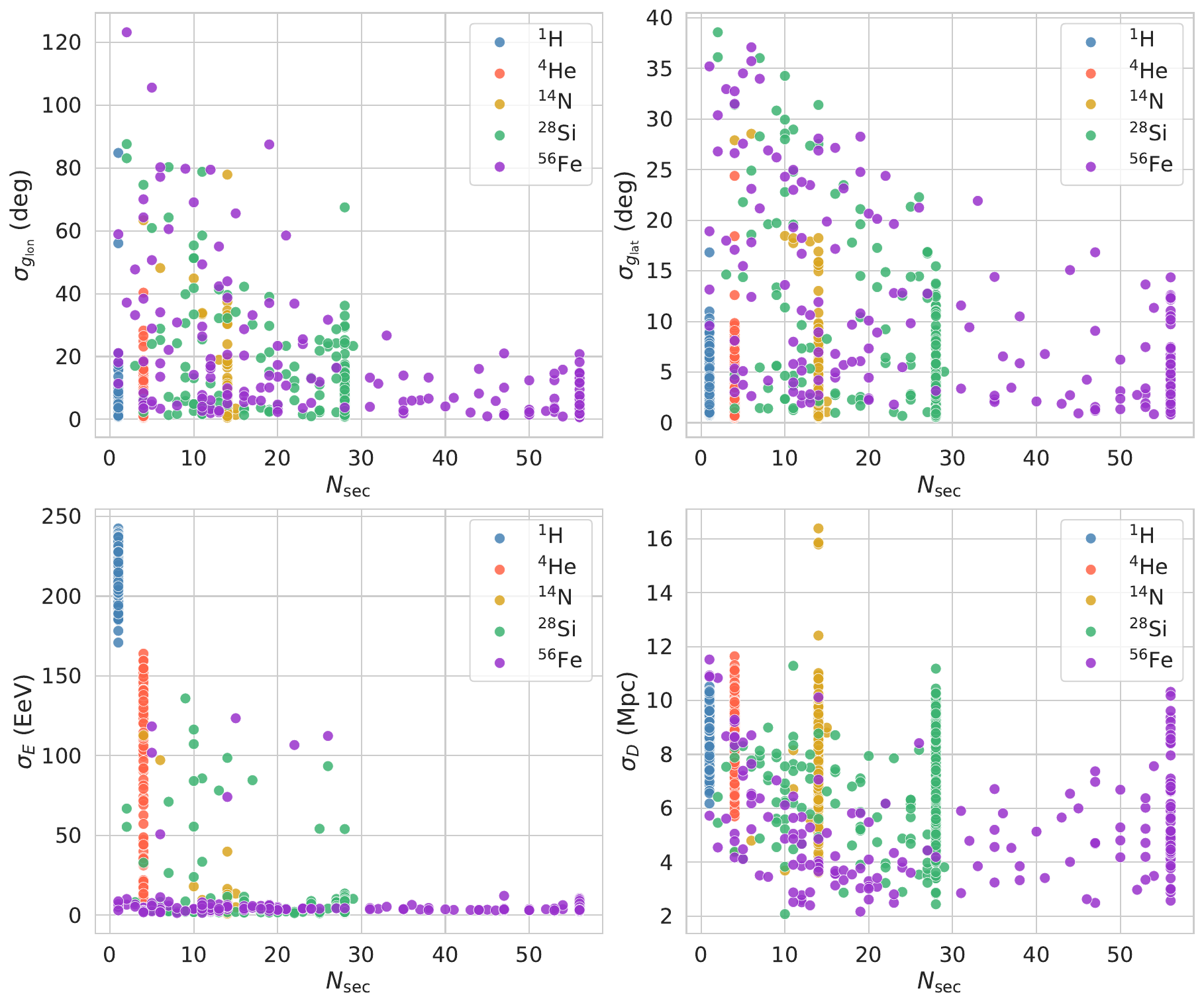}
    \caption{Posterior uncertainty \(\sigma\) for all four source parameters as a function of secondary multiplicity \(N_\mathrm{sec}\), with points colored by true primary cosmic-ray composition. Each panel corresponds to one of the four inferred source parameters: Galactic longitude \(\sigma_{g_{lon}}\), Galactic latitude \(\sigma_{g_{lat}}\), source energy \(\sigma_E\), and source distance \(\sigma_D\).}
    \label{fig:sigma_nsec}
\end{figure*}

Furthermore, Fig.~\ref{fig:sigma_nsec} shows the posterior uncertainty for all four parameters as a function of secondary multiplicity \(N_{\mathrm{sec}}\), color coded based on $\mathrm{ID}_0$ the primary cosmic ray the fragments have come from.
Events with fewer detected secondaries yield weaker directional constraints, these events tend to be the ones that start as a light primary, such as \({}^{1}\mathrm{H}\) and \({}^{4}\mathrm{He}\). However, \({}^{56}\mathrm{Fe}\) primaries can also produce large directional uncertainties when secondary multiplicity is low.

Despite this event-level scatter, the marginal prior on \(g_{\mathrm{lon}}\) is recovered with \(W_1 = 0.40^{\circ}\) (\(0.11\%\) of range), indicating that the broad individual posteriors are correctly calibrated and that no systematic directional bias accumulates across the population.

For Galactic latitude \(g_{\mathrm{lat}}\), the behavior is qualitatively similar, with posterior means that closely follow the true values throughout the full range \([-90^{\circ}, +90^{\circ}]\). The same dependence on \(N_{\mathrm{sec}}\) and \(B_{\mathrm{rms}}\sqrt{L_c}\) are present as for longitude, with uncertainty decreasing with multiplicity and increasing with field strength, and the overall scale of \(\sigma_{g_{lat}}\) is smaller.

For the source energy, \(E_{\mathrm{src}}\), the posterior means follow the one-to-one line well at low and intermediate energies. A small number of low-energy outliers show modest overestimation with large uncertainties, though these remain consistent within their credible intervals.
A subset of high-energy events appears more scattered. This is consistent with the sparsity of high-energy training examples implied by the \(E^{-1}\) injection spectrum, which naturally under-represents the high-energy tail.
For the source distance, \(D_{\mathrm{src}}\), the model shows the largest overall uncertainties of all four parameters, with \(2\sigma\) intervals that are wide across the majority of events and a visibly larger scatter around the one to one line, particularly for nearby sources.

\begin{figure*}[ht]
  \centering
  \gridline{%
    \fig{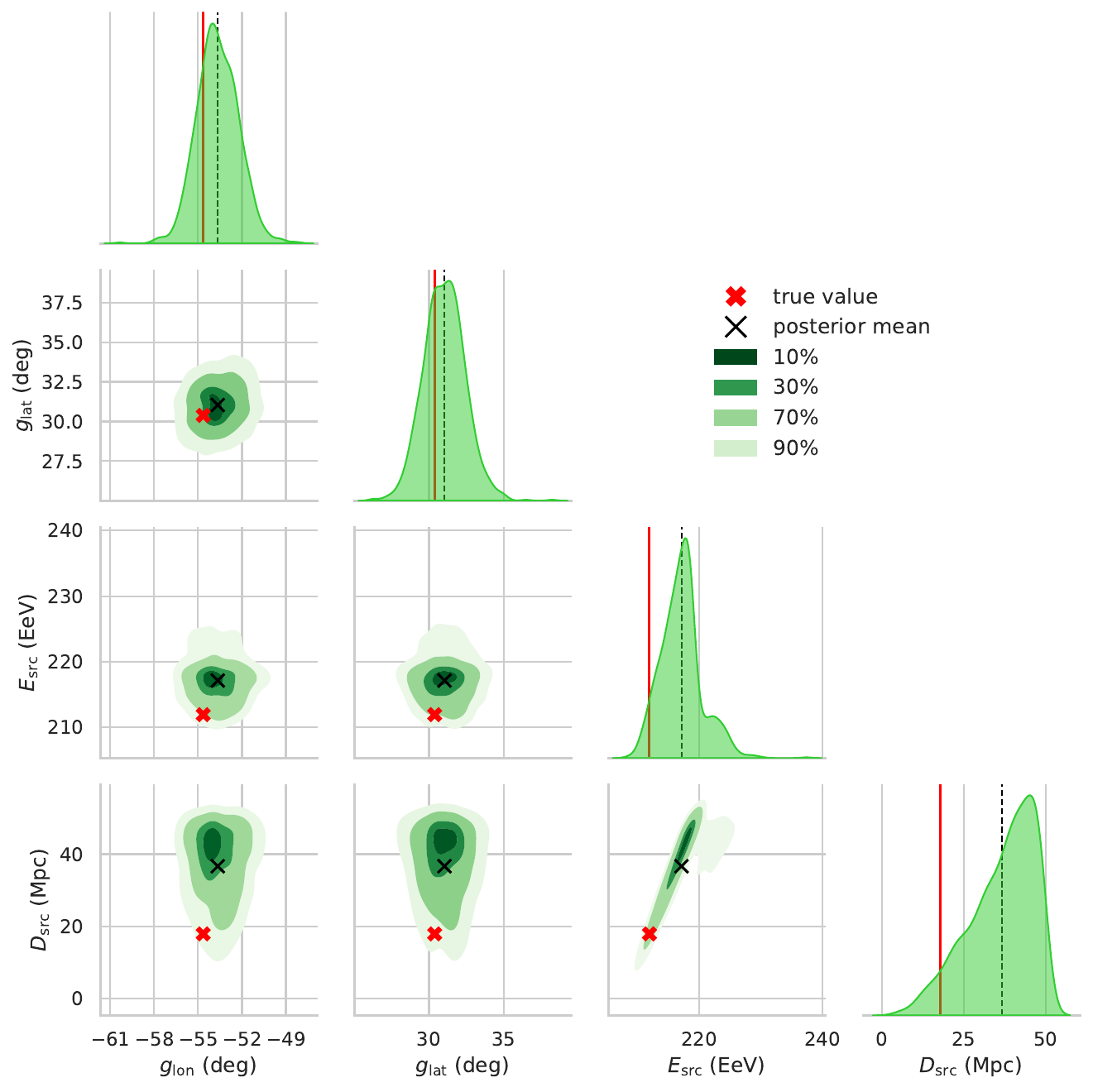}{0.52\textwidth}{(a)}%
    \fig{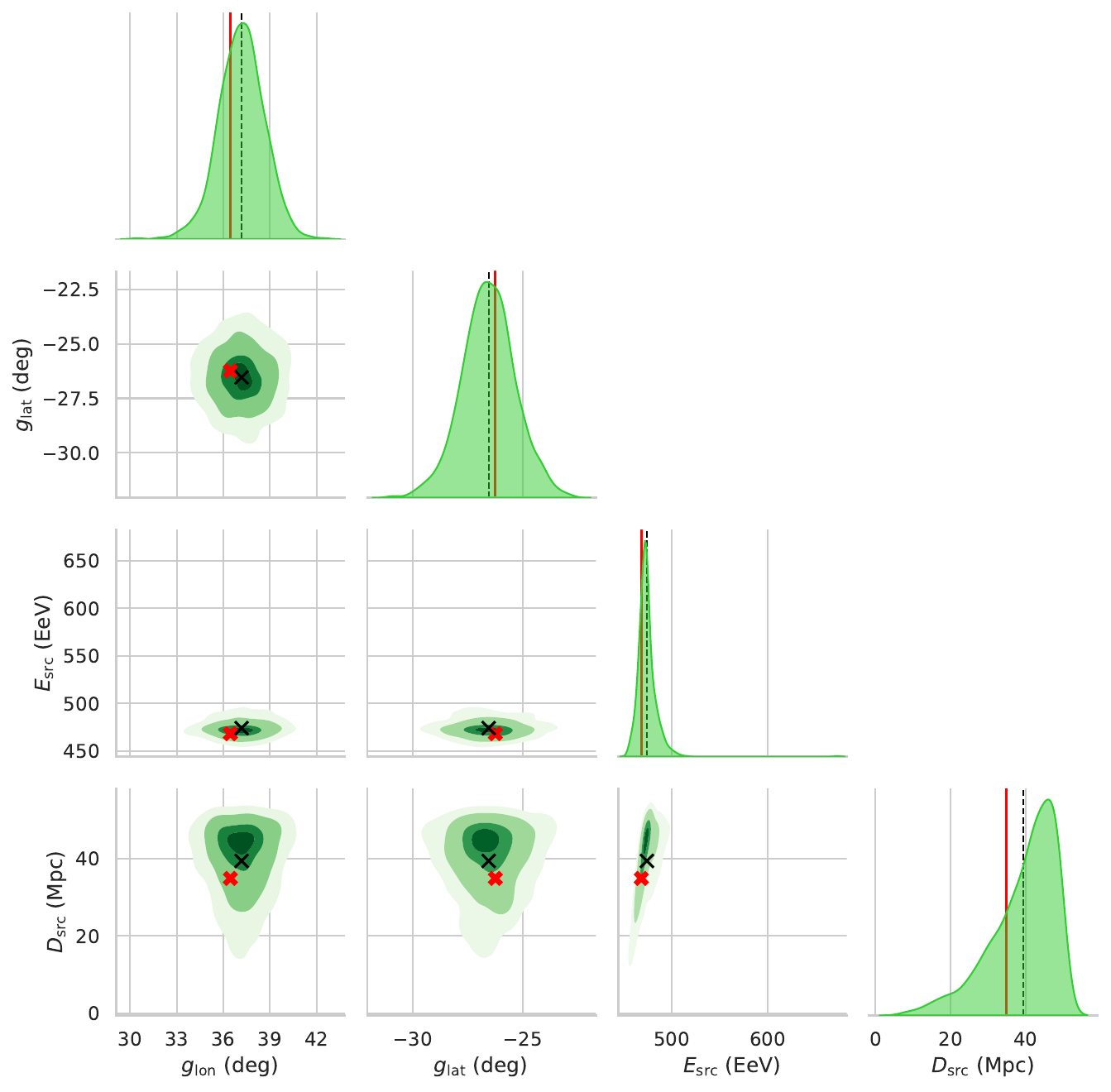}{0.52\textwidth}{(b)}%
  }
  \gridline{%
    \fig{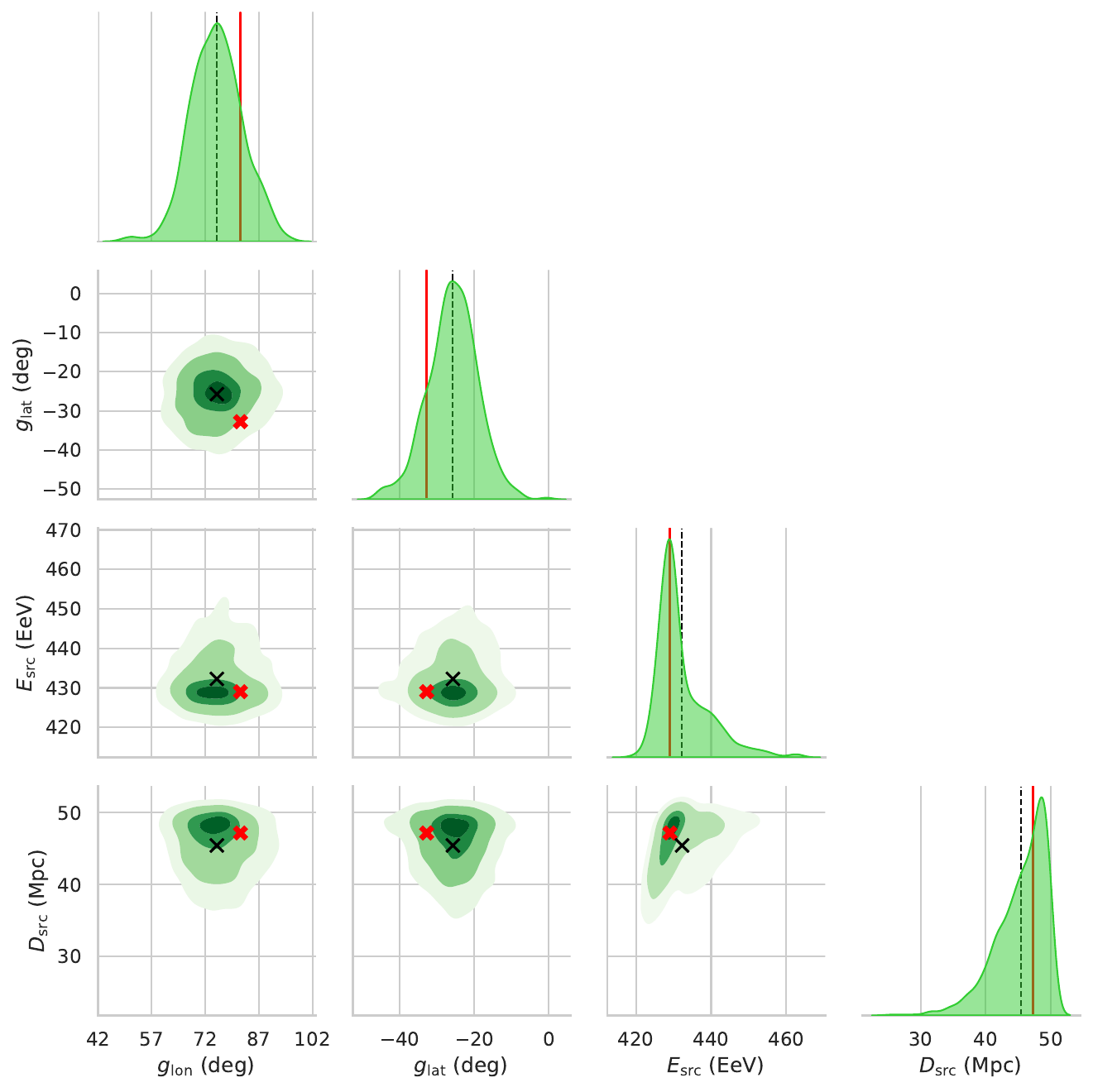}{0.52\textwidth}{(c)}%
  }
  \caption{Example posterior corner plots for three representative events. Diagonal panels show marginal posterior densities for
    $E_{\mathrm{src}}$, $D_{\mathrm{src}}$, $g_{\mathrm{lon}}$ and
    $g_{\mathrm{lat}}$. Off‑diagonal panels show two‑dimensional joint
    density contours. Red crosses mark the true parameter values; black
    crosses indicate posterior means.}
  \label{fig:corner_examples}
\end{figure*}

As a final test, in Fig.~\ref{fig:corner_examples} we show the corner plots for 3 randomly selected validation events.
Each corner plot is arranged as a 4×4 matrix: the diagonal panels display the one dimensional marginal posterior densities for $g_\mathrm{lon}$, $g_\mathrm{lat}$, $E_\mathrm{src}$  and $D_{src}$, while the off‑diagonal panels show two dimensional joint density represented by the 10\%, 30\%, 70\% and 90\% contours. The posterior mean is marked with a black cross on the joint posterior plots and a black dashed line on the marginal plots, while the true parameter values are indicated with a red cross on the joint posterior plots and a red solid line on the marginal posterior plots.

For the examples shown, the true values lie inside high density regions for most marginals and joint contours, so the posteriors are
generally well calibrated at the event level. Posterior shapes vary: some marginals are narrow and nearly symmetric, others are broad, skewed, or long tailed. This reflects differing information content in the data for each event. There are visible parameter correlations in the joint panels, for instance between $E_{\mathrm{src}}$ and $D_{\mathrm{src}}$, which the posterior captures. In these plots we can see again that $g_\mathrm{lon}$ and $g_\mathrm{lat}$ are the best constrained, since we see they have narrow posteriors and the true values are close to the posterior mean and well within the contours. The posteriors of $E_{\mathrm{src}}$ are also rather narrow, however we see that particularly for the event in panel b the true value falls within the 90\% contour. By contrast, the posterior for $D_{\mathrm{src}}$ is consistently the broadest, which again indicates that source distance is the hardest quantity for the model to recover.

Taken together, the three validation diagnostics paint a consistent picture of a well-performing model. The prior recovery test confirms that the model has learned the correct marginal structure of the parameter space, with all four source parameters recovered at the sub-percent level (\(W_1^{\mathrm{norm}} \leq 0.44\%\)), providing no evidence of systematic bias in the flow's learned representation. The scatter plots demonstrate that posterior means track the true values along the one to one relation across the full range of each parameter within the credible intervals. The corner plots further confirm that the posteriors are well calibrated at the individual event level, capturing meaningful parameter correlations such as the $E_{\mathrm{src}}$--$D_{\mathrm{src}}$ degeneracy introduced by propagation losses.

Of the four parameters, Galactic longitude is the best constrained (\(W_1^{\mathrm{norm}} = 0.11\%\)), while source distance is the most challenging, with broad posteriors that are relatively insensitive to secondary multiplicity. These limitations are physically motivated rather than indicative of model failure, and the posteriors remain centered on the true values without systematic offset. We therefore conclude that the model produces reliable, well calibrated posterior distributions suitable for inference on real data.

\section{Discussion}\label{sec:disc}

We have presented a method that uses neural posterior estimation for inferring the properties of UHECR sources from 3D \texttt{CRPropa3} simulations. The directional parameters \(g_{\mathrm{lon}}\) and \(g_{\mathrm{lat}}\) are the best constrained quantities, benefiting from the directional information encoded in the detected secondaries. The correlation between \(\sigma_{g_{lon}}\), \(\sigma_{g_{lat}}\), and \(B_{\mathrm{rms}}\sqrt{L_c}\) visible in Fig. \ref{fig:sigma_brms} is a direct reflection of the propagation physics: a stronger or more coherent EGMF deflects particle trajectories further from straight line paths, blurring the directional memory of the source. The decrease of directional uncertainty with \(N_{\mathrm{sec}}\) in Fig. \ref{fig:sigma_nsec} is also a reflection of physics phenomena, events with more detected secondaries provide more independent directional measurements that collectively point back toward the source more precisely. 

\begin{figure*}[ht]
    \centering
    \includegraphics[width=1\linewidth]{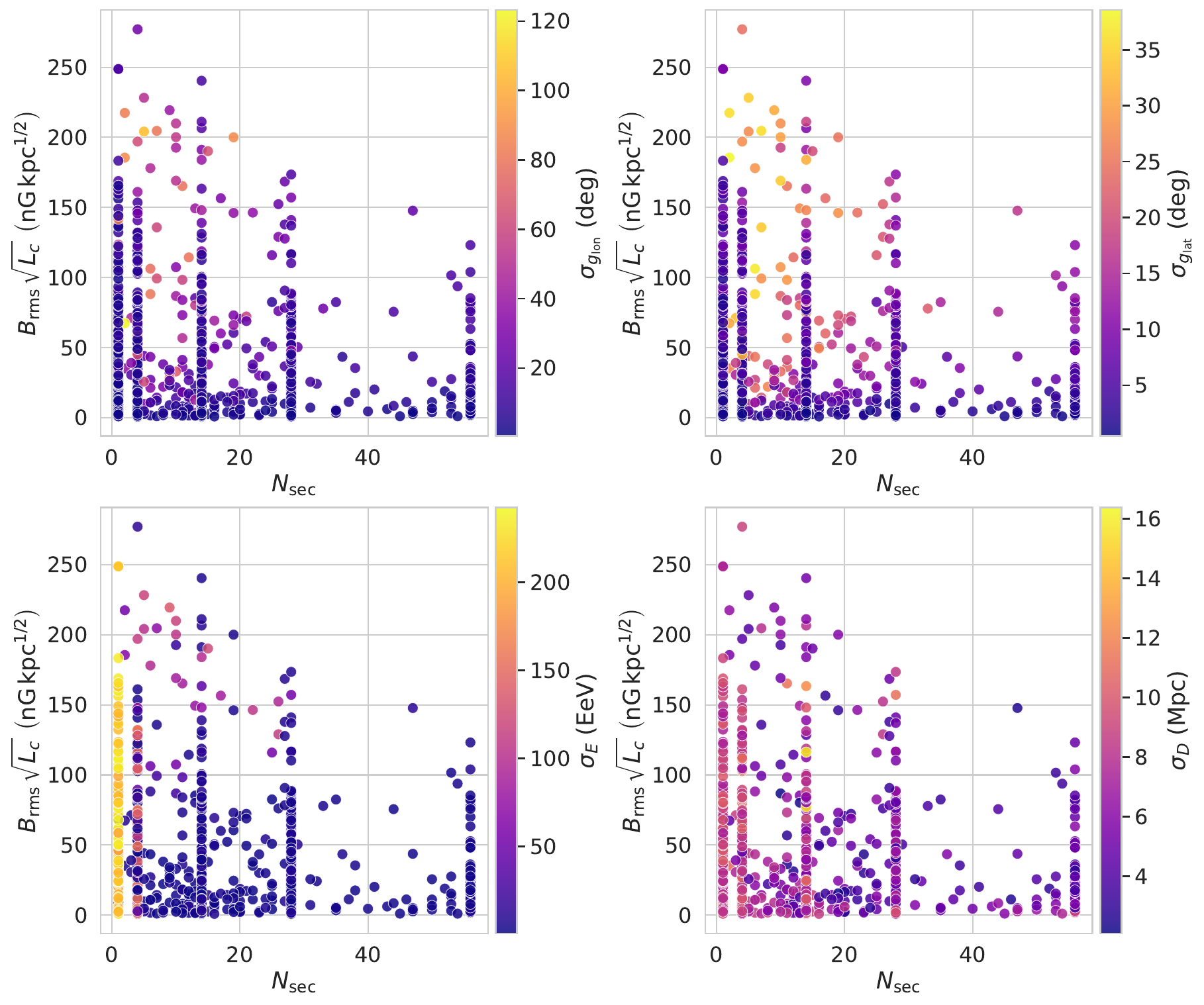}
    \caption{\(B_\mathrm{rms}\sqrt{L_c}\) for all four source parameters as a function of of secondary multiplicity \(N_\mathrm{sec}\), with points colored by posterior uncertainty \(\sigma\). Each panel corresponds to one of the four inferred source parameters: Galactic longitude \(\sigma_{g_{lon}}\), Galactic latitude \(\sigma_{g_{lat}}\), source energy \(\sigma_E\), and source distance \(\sigma_D\).}
    \label{fig:brmslc_nsec}
\end{figure*}

Fig.~\ref{fig:brmslc_nsec} illustrates how these two factors, magnetic field strength and secondary multiplicity, act jointly to determine the constraining power of each event. Each panel shows \(B_{\mathrm{rms}}\sqrt{L_c}\) as a function of \(N_{\mathrm{sec}}\), with points colored by the posterior uncertainty \(\sigma\) for each source parameter. The color gradient reveals that the best-constrained events (darkest points) cluster in the high-\(N_{\mathrm{sec}}\), low-\(B_{\mathrm{rms}}\sqrt{L_c}\) corner of each panel, while the most uncertain events (lightest points) are distributed across the low-\(N_{\mathrm{sec}}\) and high-\(B_{\mathrm{rms}}\sqrt{L_c}\) regions. This confirms that neither factor alone determines the posterior width: an event with high multiplicity but a strong EGMF can still yield broad posteriors, and conversely a low-multiplicity event in a weak field can be reasonably well constrained. The two effects are therefore complementary handles on inference quality, and both would need to be favorable for the tightest source constraints.

The significantly lower energy uncertainty for heavier primaries, visible in Fig. \ref{fig:sigma_nsec} has a clear physical explanation: photodisintegration conserves energy among detected secondaries by fragmenting the primary nucleus into lighter daughters that share the original energy, whereas photopion production transfers a significant fraction of the primary's energy to pions that decay into undetected photons and neutrinos, complicating the reconstruction of \(E_{\mathrm{src}}\).

Fig. \ref{fig:scatter} showed that in the case of \(E_{\mathrm{src}}\) there were a few outliers that are being overestimated at low energies. These may be events originating from heavy primaries where information loss due to the deactivation threshold biases the model toward overestimating the source energy, as it must reconstruct the primary energy from fewer, lighter, and more energetic fragments.

The broad posteriors for \(D_{\mathrm{src}}\) are due to the \mbox{\(p(D) \propto D^3\)} prior, which strongly weights the training distribution toward distant sources, leaving relatively fewer training examples at small distances and reducing model confidence in that regime; and \(^{56}\)Fe primaries, which carry the largest charge of all compositions considered, experience the strongest magnetic deflections, leading to longer and more curved trajectories more likely to exceed the 70\,Mpc path length threshold and be discarded from the dataset, further reducing the available information for heavy primary events. Despite these limitations, the distance posteriors remain centered on the true values without systematic offset, confirming that the broad uncertainties are physically motivated rather than indicative of model failure. The near perfect primary composition classification, with misclassification confined exclusively to adjacent mass groups, is consistent with the charge-dependent similarity of propagation signatures between neighboring nuclei. This result demonstrates that the Deep Set compression architecture effectively encodes compositional information from the secondary particle set, and suggests that the framework will be well positioned to exploit the improved composition sensitivity of the AugerPrime upgrade \citep{Castellina:2019jd}.

Having validated the model's ability to infer source properties and composition from extragalactic propagation alone, we now turn to the question of Galactic propagation. In order to apply our model to UHECR measurements we would need to backtrack the arrival directions through a model for the GMF. We made this choice to adopt a shell geometry combined with backtracking, rather than a full forward simulation including GMF lensing to improve computational efficiency. The shell approach allows us to capture all simulated events and change the GMF model a posteriori without retraining the neural posterior estimator. While this approximation does not capture the full forward model of UHECR detection, requiring a simplified treatment of detector and magnetic field uncertainties, it provides a flexible and tractable framework for inference at the highest energies. In contrast, a full forward model with Galactic lensing would naturally incorporate these uncertainties and enable conditioning on different GMF realizations, as well as the exploration of more realistic EGMF scenarios in future. However, such an approach is significantly more computationally demanding, requiring substantially larger training datasets and careful tuning of simulation and targeting strategies to achieve sufficient coverage of the relevant parameter space.

Finally, while our presented approach uses NPE, there are other ways density estimators can enter the analysis pipeline. In neural likelihood estimation the neural network learns $p(x | \theta)$, which is combined with a prior for $p(\theta)$ to sample from the posterior (e.g, used to infer neutron star parameters in \citealt{Brandes_2024}). A benefit for neural likelihood estimation is it can be used to do inference with different priors than were used for training, since $p(x | \theta)$ a neural surrogate for the simulator. However, to generate samples from the posterior, a sampler still needs to be run, as was done in the ABC method. Since the normalizing flow is cheap to sample from, the samples will be much quicker to generate that running \texttt{CRPropa 3} in the inference loop, as was done with the ABC method. Furthermore, $\nabla_\theta p(x | \theta)$ can also be easily computed and so gradient-enhanced samplers such as Hamiltonian MC could also be used to optimize the sampler's exploration of the posterior.
In our setup, neural likelihood estimation would have also required generating a higher dimensional $x$ of varying cardinality, a more challenging modelling task than the inverse direction.
For these reasons, we haven't explored neural likelihood estimation here, and leave these studies of its efficacy for UHECRs to future work.

\section{Conclusion}\label{sec:conclusion}

We have presented a simulation-based inference framework that learns the posterior distribution for ultra-high energy cosmic ray source properties directly from three-dimensional \texttt{CRPropa~3} propagation simulations. By combining a Deep Set encoder to handle variable length secondary particle sets with a neural spline flow for density estimation, the model produces calibrated, amortized posteriors for source direction, energy, distance, and primary composition for individual events.

Validation on held-out simulations demonstrates that the framework recovers all source parameters without systematic bias. Directional parameters are the best constrained, reflecting the directional information encoded in secondary momenta. The source distance remains the most uncertain parameter, a limitation that is physical rather than methodological: the \(p(D)\propto D^{3}\) prior weights the training distribution toward distant sources, and magnetic deflections increasingly erase distance information for heavy primaries. The auxiliary classification head achieves high accuracy across all mass groups (\(\geq 98.2\%\)), with misclassifications confined to adjacent compositions, consistent with the charge-dependent similarity of propagation signatures.

The model's conditional design treating the EGMF parameters \(B_{\mathrm{rms}}\) and \(L_{c}\) as inputs rather than inference targets is a pragmatic choice that enables stable training with approximately 5 million events. At test time, this allows marginalization over EGMF uncertainties via ensemble averaging across plausible field realizations, or fixed field inference when a specific field model is assumed. Future work will extend the framework to jointly infer EGMF properties, incorporate Galactic magnetic field lensing directly rather than via post-hoc backtracking, and scale to larger UHECR datasets expected from AugerPrime and TAX4 \citep{Castellina:2019jd, Abbasi:2021pd}.

More broadly, this work demonstrates that neural posterior estimation provides a scalable and interpretable inference framework for UHECR astrophysics. This framework is readily adaptable to different source scenarios, magnetic field models, and observables, offering a flexible tool for source identification as the era of high statistics, high composition precision UHECR data approaches.

\begin{acknowledgments}

We thank L.Heinrich and A. Kofler for their valuable input during our discussions.

N. Bourriche and N. Hartman acknowledge the financial support from the Excellence Cluster ORIGINS, which is funded by the Deutsche Forschungsgemeinschaft (DFG, German Research Foundation) under Germany’s Excellence Strategy - EXC-2094-390783311.

\end{acknowledgments}

Software: \texttt{pytorch} \citep{Paszke2019PyTorch}, \texttt{nflows} \citep{durkan2019neuralsplineflows}, \texttt{CRPropa3} \citep{Batista:2022pd}

\bibliography{main}{}
\bibliographystyle{aasjournal}

\end{document}